
\input phyzzx
\def\winf{$W_\infty$}

\def\psid{\Psi^\dagger}
\def\del{\partial}
\def\tr{{\rm tr}}

\def\a{\bar A}
\def\ket{\rangle}
\def\bra{\langle}
\def\e{\epsilon}
\def\half{{1\over 2}}

\indent \hfill{TIFR-TH-92/40}\break
\indent \hfill June, 1992 \break
\date{}
\titlepage
\title{\bf Non-relativistic Fermions, Coadjoint Orbits of \winf\ and
String Field Theory at $c=1$}
\author{Avinash Dhar\foot{adhar@tifrvax.bitnet.},
Gautam Mandal\foot{mandal@tifrvax.bitnet.}
and Spenta R. Wadia
\foot{wadia@tifrvax.bitnet}}
\address{ Tata Institute of Fundamental Research, Homi Bhabha Road, Bombay
400 005, India}
\abstract{We apply the method of coadjoint orbits of  \winf-algebra to
the problem of non-relativistic fermions in one dimension. This leads to a
geometric formulation of the quantum theory in terms of the quantum phase
space distribution of the fermi fluid. The action has an infinite series
expansion in the string coupling, which to leading order reduces to the
previously discussed geometric action for the classical fermi fluid based
on the group $w_\infty$ of area-preserving diffeomorphisms. We briefly
discuss the strong coupling limit of the string theory which, unlike the
weak coupling regime, does not seem to admit of a two dimensional
space-time picture. Our methods are equally applicable to interacting
fermions in one dimension.}
\endpage

\section{\bf Introduction:}

Non-relativistic fermions in one dimension have recently been investigated
in connection with models of two-dimensional string theory. The connection
proceeds by realizing that two-dimensional string theory (in flat spacetime
and linear dilaton background) is perturbatively
equivalent to two-dimensional
\REF\DNW{S. Das, S. Naik and S.R. Wadia, Mod. Phys. Lett. A4 (1989) 1033.}
\REF\DJNW{A. Dhar, T. Jayaraman, K.S. Narain and
S.R. Wadia, Mod. Phys. Lett. A5 (1990) 863.}
\REF\DDW{S. Das, A. Dhar and S.R. Wadia, Mod. Phys. Lett. A5 (1990) 799.}
\REF\POLA{J. Polchinski, Nucl. Phys. B234 (1989) 123.}
\REF\BL{T. Banks and J. Lykken, Nucl. Phys. B331 (1990) 173.}
``Liouville gravity'' coupled to one-dimensional matter [\DNW-\BL].
The lattice
formulation of the latter is described by a hermitian matrix model in one
dimension
\REF\KAZ{V.A. Kazakov and A.A. Migdal, Nucl. Phys. B320 (1989) 654.}
[\KAZ],
which in turn is exactly mapped onto a theory of
nonrelativistic fermions in one dimension
\REF\BIPZ{E. Brezin, C. Itzykson, G. Parisi
and J.B. Zuber, Comm. Math. Phys. 59 (1978) 35.}
[\BIPZ].
There are various reasons why
it is of interest to write down an exact bosonization of this model. It
would provide us with  an exact field theory action of two-dimensional
string theory, with manifest invariance principles.
It would make possible a description of stringy non-perturbative
behaviour. It would also be of
interest from the viewpoint of condensed matter physics, where this
problem was first posed and approximately solved by Tomonaga
\REF\TOMONAGA{S. Tomonaga, Prog. Theor. Phys. 5 (1950) 544.}
[\TOMONAGA].
An exactly
solvable version of Tomonaga's model was  formulated by Luttinger
\REF\LUTTINGER{J.M. Luttinger, J. Math Phys. 4 (1963) 1154.}
[\LUTTINGER] where
fermions  obey the relativistic dispersion relation $ E(k) = \pm |k|$.
This version of the model has been the subject of much  study and
elaboration
\REF\LM{Lieb and  Mattis, J. Math. Phys. 6 (1965) 304.}
\REF\HALDANE{F.D.M. Haldane, J. Phys. C14 (1981) 2585.}
[\LM,\HALDANE].
There are also connections with quantum Hall effect in two
dimensions. In a completely different area of activity connected with the
study of the large $N$ limit of matrix models, this problem was studied by
the collective field formulation
\REF\JS{A. Jevicki and
B. Sakita, Nucl. Phys. B165 (1980) 511.}\REF\DJ{S.R. Das and A.
Jevicki, Mod. Phys. Lett. A5 (1990) 1639.}
[\JS], which was also adopted for the study of
two-dimensional string field theory
[\DJ]. A perturbative (low energy) expansion and a treatment of the
turning point problem was given in
\REF\SW{A.M. Sengupta and S.R. Wadia, Int. J. Mod. Phys. A6 (1991) 1961;
G. Mandal, A.M. Sengupta and S.R. Wadia, Mod. Phys. Lett. A6 (1991) 1465.}
[\SW].
In a series of papers we have studied
the nonrelativistic fermion problem from the viewpoint of $W_\infty$
symmetry
and its classical limit $w_\infty$ ($= $ the area-preserving
diffeomorphisms in two dimensions)
\REF\DDMWA{S. Das, A. Dhar, G. Mandal and S.R. Wadia, ETH, IAS and Tata
preprint, ETH-TH-91/30, IASSNS-HEP-91/52 and TIFR-TH-91/44 (Sept. 1991),
to appear in Int. J. Mod. Phys. A (1992).}
\REF\DDMWB{S. Das, A. Dhar, G. Mandal and S.R. Wadia, Mod. Phys. Lett. A7
(1992) 71.}
\REF\DDMWC{S. Das, A. Dhar, G. Mandal and S.R. Wadia,
Mod. Phys. Lett. A7 (1992) 937.}
\REF\DMW{A. Dhar, G. Mandal and S.R. Wadia, IAS and Tata preprint,
IASSNS-HEP-91/89 and TIFR-TH-91/61, to appear in Int. J. Mod. Phys. A
(1992).}
[\DDMWA-\DMW]. See also
\REF\DANIELSSON{U.H. Danielsson, {\sl A Study of Two Dimensional String
Theory}, Ph. D. Dissertation, Princeton University, 1992.}
[\DANIELSSON].
In [\DMW] we discussed the
classical limit as an incompressible fermi fluid in two dimensions. Using
the method of the co-adjoint orbits of $w_\infty$ we presented a
geometrical action and a string picture in terms of the classical phase
space of the fermi fluid.  We have also made precise statements about the
limitations of the collective field method.
In this paper we extend the results of [\DMW]
and  give an exact discussion of the
bosonization, using the method of coadjoint orbits of the group
$W_\infty$. The nonrelativistic fermions give rise to specific coadjoint
orbits of $W_\infty$. These coadjoint orbits are specified by a quadratic
constraint and by the number
of particles or equivalently by the fermi level. There is a close analogy
with the problem of an $SU(2)$  spin in a magnetic field, where the
coadjoint orbits of $SU(2)$ are specified by the values of the total
angular momentum. In case the spin is formed out of a two-state fermi
system, the coadjoint orbit corresponds to spin half if the number
of fermions is one (half filled) and to spin zero if the number of
fermions is zero (unfilled) or two (completely filled). In exact analogy
with the spin problem we present an action functional on those coadjoint
orbits of $W_\infty$ (specified by an appropriate set of constraints)
which correspond to non-relativistic fermions in one
dimension. This action is manifestly invariant under the
$W_\infty$ transformations that are a symmetry of the original fermionic
action. We emphasize that the symmetry group is $W_\infty$ and not
$w_\infty$. The latter is obtained only in the limit $\hbar$ (string
coupling) $\to 0$. There is a way of writing the action in terms of
a scalar field in $2+1$ dimensions. This field can be interpreted
as the ``phase space'' distribution function of the original
one-dimensional fermi fluid. A novel feature of the action is that
it can be formally expanded in an
infinite series in $\hbar$ or the string coupling. The leading term
reduces to the geometric action presented in [\DMW], which is based on
$w_\infty$ symmetry. In the strong coupling limit, $\hbar \to \infty$,
however, this picture clearly breaks down. Indeed, it seems that an
interpretation in terms of a two-dimensional target space theory does not
exist. This seems to suggest that the standard reasoning that the
dynamical metric on the world-sheet is equivalent to one conformal
(Liouville) mode which in turn gives rise to one additional target space
dimension does not work in the strong coupling limit.

The bosonization technique we have developed here is also applicable to
the case of interacting fermions in one dimension.

The plan of the paper is as follows. In the next section we review some
aspects of the formulation of fermion field theory and \winf\
algebra as developed in
[\DDMWA-\DMW]. This will also serve to set up our notation. In Sec. 3 we
discuss in detail the analogy of the present problem with that of a spin
in a magnetic field. Indeed, the problems are identical, except that the
``rotation'' group in the present case is \winf. We show that the
bilocal operator, which is the analogue of the spin operator in the
present case, satisfies a constraint that determines the representation to
which the \winf\ spin belongs, analogous to the constraint $\vec
S^2=$ constant\ \  for the
rotation group which determines the spin content.
We write down the ``classical'' bosonized action in Sec. 4, in exact
analogy with that for a spin in a magnetic field. The group for which the
action is written down is \winf, which is a one-parameter
deformation of $w_\infty$, the group of area-preserving
diffeomorphisms in two dimensions. The parameter is $\hbar$ and in the
present case is identified with the string coupling. The ``classical''
action may, therefore, be thought of as an infinite series in string
coupling.  In Sec. 5 we discuss solutions  to the classical equation of
motion which satisfy the constraints on the bilocal operator.
The constraints can be solved only perturbatively in $\hbar$,
the string coupling constant.
We  show that at the lowest order in $\hbar$
the solutions are characteristic
functions, as one might expect for a classical fermi fluid. In Sec. 6 we
discuss  how in the
$\hbar\to 0$ limit the results of [\DMW] are reproduced.
In Sec. 7 we indicate how \underbar{interacting} fermions in one dimension
can be treated by our bosonization technique.
Finally, in Sec. 8 we end with some  concluding remarks.

\section{\bf Fermion Field Theory and $W$-infinity algebra:}

In the gauge theory formulation of [\DDMWA-\DDMWB], the action for the
fermion field  theory which is equivalent to the $c=1$ matrix model, is
$$S[\Psi, \psid, \a]= \int dt\, \bra \Psi(t) | (i\hbar \del_t + \a(t))|
\Psi(t) \ket \eqn\twoone$$
where $\a(t)$ is some given background field. The fermion field $|
\Psi(t)\ket $ is a ket vector in the single-particle Hilbert space with
components  $\bra x | \Psi(t) \ket \equiv
\psi(x,t)$ in the coordinate basis.
In the same basis, the matrix elements of $\a(t)$ will be
denoted by $\bra x | \a(t) | y \ket \equiv  \a(x,y,t) $. For the $c=1$
matrix model,
$$\a(x,y,t)= \half(\hbar^2 \del_x^2 - V(x)) \delta(x-y),\quad V(x) = -x^2
+ {g_3 \over \sqrt{N}} x^3 + \cdots   \eqn\twoonea$$
In writing \twoone-\twoonea\ we have chosen the zeroes of the energy and
$x$-axis appropriately such that the
(quadratic) maximum of the potential occurs at $x=0$ and $V_{\rm max} =
V(0) =0$. We have also introduced appropriate rescalings
suitable for the double scaling limit. The parameter $N$ that appears in
\twoonea\ is the total number of fermions,
$$N= \bra \Psi(t) | \Psi(t) \ket = \int dx\, \psi^\dagger(x,t) \psi(x,t)
\eqn\twooneb$$
which is taken to infinity in the double scaling limit. The other
parameter that appears in \twoone\
and \twoonea, i.e. $\hbar$, is the string
coupling constant (see {\it e.g}
\REF\POLB{J. Polchinski, Nucl. Phys. B346 (1990) 253.}
[\POLB]). The quantum
theory is defined by the functional integral
$$Z =\int {\cal D}\Psi, {\cal D}\Psi^\dagger
\exp {i\over \hbar}S(\Psi,\Psi^\dagger,\a)  \eqn\twoonec$$

The action \twoone\ has the background gauge invariance
$$ \eqalign{
|\Psi(t) \ket  &\to V(t) |\Psi(t) \ket \cr
\a(t) &\to V(t) \a(t) V^\dagger(t) + i \hbar V(t) \del_t V^\dagger (t)
} \eqn\twotwo $$
where $V(t)$ is a unitary operator in the single-particle Hilbert space.
For a given fixed $\a(t)$, the residual gauge symmetry is determined by
$$i\hbar \del_t V(t) + [\a(t), V(t)]=0 \eqn\twothree $$
with the solution
$$ V(t) = {\cal U}(t) V_0 {\cal U}^\dagger(t),
\quad {\cal U}(t) = {\cal P}
\exp[ {i\over \hbar} \int^t d\tau\, \a(\tau)]. \eqn\twofour$$
Thus the residual symmetry, for any given $\a(t)$, is parametrized by an
arbitrary constant unitary operator $V_0$. The set of all the $V_0$'s
forms the group \winf.

The \winf\ algebra is the algebra of differential operators in the
single-particle  Hilbert space
\REF\POPE{C.N.Pope, L.J.Romans and X.Shen, ``A brief History of
$W_\infty$,'' in {\sl Strings 90}, ed. R.Arnowitt et al (World
Scientific, 1991), and references therein.}
\REF\BAKAS{I. Bakas and E.B. Kiritsis, Int. J. Mod. Phys. A6 (1991) 2871.}
[\POPE,\BAKAS].
A convenient way to describe it is by introducing the generating function,
$$ \hat g(\alpha, \beta) \equiv \exp i(\alpha \hat x- \beta \hat p), \quad
[\hat x, \hat p]= i\hbar \eqn\twofive$$
The product law
$$ \hat g(\alpha, \beta) \hat g(\alpha', \beta') = \exp [{i\hbar \over 2}
(\alpha\beta' -\alpha' \beta)] \hat g(\alpha+ \alpha', \beta+ \beta').
\eqn\twosix$$
is a well-known consequence of the Heisenberg algebra.
The \winf\ algebra is a straightforward  consequence of \twosix: $$ [\hat
g(\alpha, \beta), \hat g(\alpha', \beta')] = 2i\sin [{\hbar\over 2} (\alpha
\beta' -\alpha' \beta)] \hat g(\alpha+ \alpha', \beta+ \beta').
\eqn\twoseven$$
The $\hat g(\alpha, \beta)$ form an ``orthogonal'' basis for the \winf\
algebra. That is,
$$\tr [\hat g(\alpha,\beta) \hat g(\alpha',\beta')]= {2\pi \over \hbar}
\delta (\alpha + \alpha') \delta(\beta + \beta') \eqn\twoeight$$
This can be easily proved, for example by evaluating the trace in the
coordinate basis and by using the fact that the matrix elements of $\hat
g(\alpha, \beta)$ are
$$\bra x | \hat g(\alpha ,\beta) | y \ket = \delta(x-y + \hbar \beta)
\exp(i\alpha {x+y \over 2}) \eqn\twonine$$
The notation `tr' in \twoeight\ stands for integration over $x,y$
etc.

A general element $\Theta$ of \winf\ algebra may, therefore, be written as
$$\Theta = \int d\alpha\, d\beta\, \theta(\alpha, \beta) \hat g(\alpha,
\beta)  \eqn\twoten $$
Since $\hat g(\alpha, \beta)$ satisfies the hermiticity condition $\hat
g(\alpha, \beta) = \hat g(-\alpha, -\beta)$, we see from \twoten\ that for
hermitian $\Theta$ we must have $\theta^*(\alpha, \beta) = \theta(-\alpha,
-\beta)$. Because of this hermiticity condition $\theta(\alpha, \beta)$
can be expressed in terms of a real function $u(p,q)$:
$$\theta (\alpha, \beta) = \int {dp\over 2\pi} {dq\over 2\pi} u(p,q) \exp
i(p\beta- q\alpha)  \eqn\twoeleven$$
Equations \twoten\ and \twoeleven\ define the Weyl correspondence between
functions in phase space ($u(p,q)$) and operators ($\Theta$).
As we shall see later, the functions $u(p,q)$ will later turn out to be
closely related to the phase space density of the fermion theory.

The unitary operators $V_0$ appearing in \twofour\ may now be constructed
by exponentiating the general element of the \winf\ algebra in \twoten. To
end this section we note that the algebra in \twoseven\ reduces to the
algebra of area-preserving diffeomorphisms in two dimensions in the limit
$\hbar\to 0$. The \winf\ group that we are dealing with therefore is a
quantum deformation of the group of area-preserving diffeomorphisms in two
dimensions, the parameter of deformation being $\hbar$ or the string
coupling.

\section{\bf The Bilocal Operator, the Constraint and Analogy with
Spin in a Magnetic Field:}

The analogy between the present problem and that of a spin in a magnetic
field has already been pointed out by us in [\DDMWB] and [\DMW].
In this section we
will elaborate on that analogy further and show that, in fact, the two
problems are closely related. The ``rotation'' group in this case is
\winf.

The appropriate ``spin'' variable in the present context is
[\DDMWA-\DDMWC] the fermion bilocal operator $\Phi(t)$ defined as follows
$$\Phi(t) \equiv | \Psi(t) \ket \bra \Psi(t) |   \eqn\threeone$$
In the coordinate basis, the $xy$-component is given by
$$\Phi(x,y,t)\equiv \bra x | \Phi(t) | y\ket  = \Psi(x,t) \psid(y,t)
\eqn\threetwo$$
Under  \winf\ ``rotations'' of the fermion field, the bilocal operator,
which is gauge-covariant by construction,
transforms by the  adjoint action of the group:
$$ | \Psi(t) \ket  \rightarrow  V|\Psi(t) \ket \quad\Rightarrow \Phi(t)
\rightarrow V \Phi(t) V^\dagger  \eqn\threethree$$
We may expand $\Phi(t)$ in the basis $\hat g(\alpha, \beta)$ provided by
the Heisenberg-Weyl group. We have,
$$ \Phi(t) = {\hbar \over 2\pi} \int d\alpha \, d\beta\, W(\alpha, \beta,
t) \hat g(\alpha, \beta)    \eqn\threefour $$
where the fermion bilocal operator
$$ W(\alpha, \beta, t) \equiv \int dx\, \psi(x+ \half \hbar \beta, t)
\psi^\dagger
(x- \half \hbar \beta, t) \exp (i\alpha x)    \eqn\threefive$$
provides a field theoretic representation of \winf\ algebra:
$$[ W(\alpha, \beta, t), W(\alpha', \beta',t) ]= 2i \sin [{\hbar \over 2}
(\alpha\beta' - \alpha' \beta)] W(\alpha + \alpha', \beta + \beta', t)
\eqn\threesix $$
Finally, using the equation of motion for the fermion ket $| \Psi(t)
\ket$, which can be obtained by varying action \twoone, one can easily
obtain the equation of motion for $\Phi(t)$:
$$ i\hbar \del_t \Phi(t) + [\a(t), \Phi(t)]=0. \eqn\threeseven$$

Equations \threefour, \threesix\ and \threeseven\ are exactly like the
corresponding equations for a spin in a magnetic field. Let $S^i(t)$ be
the spin variable, $T^i$ the generators of $SU(2)$ (in the appropriate
representation). Then, the operator $S(t)= \sum_i S^i(t) T^i$ is like
$\Phi(t)$, $S^i(t)$ being like $W(\alpha, \beta, t)$ and $T^i$ like $\hat
g(\alpha, \beta)$. The algebra of $W(\alpha, \beta,t)$'s is like the spin
algebra $[S^i(t), S^j(t)] = i \e^{ijk} S^k(t)$. The equation of motion
$\del_t S^i(t)= (1/i)[\vec B.\vec S, S^i]= \e^{ijk} B^j S^k(t) $ can be
rewritten in terms of $S(t)$ and $B\equiv \sum_i B^i T^i$ and reads
$i\del_t S(t) = [B, S(t)]$, which is like \threeseven with $B$ playing the
role of $-\a$. The analogy between
the two cases is therefore complete. In the case of the $SU(2)$ spin the
problem is completely specified by further specifying the representation
to which the spin belongs. This may be done, for example, by specifying
the value of $\sum_i [S^i(t)]^2$. This is equivalent to giving a quadratic
equation for the matrix $S(t)$, as may be easily verified. Another  way of
specifying the representation to which the spin belongs is by giving an
explicit representation for $S^i(t)$ in terms of more elementary objects.
For example, the spin $1/2$ (and spin $0$) representation can be
constructed in terms of a spin-$1/2$ fermi system. Let us study this
representation in more detail since this is what happens for the \winf\
spin that is of interest to us in this work.

Let us assume that the spin variable $S^i(t)$ has a more microscopic
representation in terms of spin $1/2$ fermions  $\psi_a(t)\, (a=1,2)$:
$$ S^i(t) = \psi^\dagger(t) {\sigma^i\over 2} \psi(t) \eqn\threeeight$$
where $\sigma^i$ are the Pauli matrices satisfying
$$\eqalign{
[&{\sigma^i \over 2}, {\sigma^j\over 2}] = i \e^{ijk} {\sigma^k \over 2}\cr
\{&{\sigma^i \over 2}, {\sigma^j\over 2}\} = \half \delta^{ij}\cr}
\eqn\threenine$$
Using equal-time fermion anticommunication relations it is easy to verify
that \threeeight\ satisfies $[S^i(t), S^j(t)] = i \e^{ijk} S^k(t)$.
Further, it can be easily verified, using the identity
$$ \sum_i ({\sigma^i \over 2})_{ab}({\sigma^i \over 2})_{a'b'} = \half
(\delta_{ab'} \delta_{a'b} - \half \delta_{ab} \delta_{a'b'})
\eqn\threeten$$
that the $S^i(t)$ are characterized by the relation
$$ \sum_i [S^i(t)]^2 = {3\over 4} n_f (2-n_f)   \eqn\threeeleven$$
where $n_f= \sum_a \psi^\dagger_a(t) \psi_a(t)$ is a Casimir operator
(since it commutes with all $S^i(t)$). It  simply measures the total
number of filled levels in any state, which is a fixed number for all the
states of the system and equals the total number of fermions. So, in this
simple case of a two-level system we are led to the constraint
\threeeleven. For half-filling, $n_f=1$, we find $[S^i(t)]^2 = 3/4$, which
is the correct value of the Casimir for spin $1/2$. This accords with the
fact that in this case of a two-level system, half-filling corresponds
to a two-state system---one, the  fermi vacuum in which the lower of the
two states is occupied, and the other one is the excited state in which
the fermion in the vacuum is excited to the higher level. For no filling
(or equivalently complete filling) there is only the fermi vacuum and no
excited states. Therefore, $\sum_i [S^i(t)]^2=0$ is appropriate for this
case. So, we see that information regarding which spin representation the
system belongs to is contained in the constraint \threeeleven\ (which
follows from the representation \threeeight) and depends only on the
filling of the fermi sea. An identical situation arises in our present
case of interest of \winf\ spin. Before we discuss that, note that in the
$SU(2)$ case the constraint \threeeleven\ is equivalent to a quadratic
equation for $S(t)$. In fact, one can show that $4 S(t)^2 + S(t)= \sum_i
[S^i(t)]^2$. We have mentioned this because there are an infinite number
of Casimirs for \winf. Since the above type of quadratic equation contains
information about all of them, it is easier to deduce this type of
relation in this case.

The above line of argument can be applied identically to the present case
of \winf\ spin. The ``spin variable'' $\Phi(t)$ has a microscopic
representation in terms of fermions \threeone. Thus,
$$\eqalign{
\bra x| [\Phi(t)]^2 | y \ket =& \int dz \bra x | \Phi(t) | z \ket\bra z |
\Phi(t) | y \ket \cr
=& \int dz\,\psi(x,t) \psi^\dagger (z,t) \psi(z,t) \psi^\dagger (y,t) \cr
=&\; \bra x | \Phi(t) | y \ket (1+ N),\cr}$$
i.e.
$$(\Phi(t))^2 = (1+ N) \Phi(t)
 \eqn\threetwelve $$
where now $N = \int dx\, \psi^\dagger (x,t) \psi(x,t) $ and is again a
Casimir since it commutes with all $W(\alpha, \beta, t)$. It is the total
number of filled levels in any state, that is, the total number of
fermions. The constraint \threetwelve, together with the constraint on
total number of fermions, fixes the representation to which the
$W_\infty$-spin $\Phi(t)$ belongs.

\section{\bf The Action}

The most elegant way of arriving at an action for this problem is to
follow Kirillov's method of coadjoint orbits
\REF\KIRILLOV{See for example,
A.A. Kirillov, {\sl Elements of the Theory of Representations} (1976);
A. Alekseev and S. Shatasvili, Nucl. Phys. B 323 (1989) 719;
A. Alekseev, L. Faddeev and S. Shatasvili, J. Geom. Phys. 1 (1989) 3;
B. Rai and V.G.J. Rogers, Nucl. Phys. B341 (1990) 119.}
[\KIRILLOV].
We will briefly outline the
procedure first for the case of spin in a magnetic field. (For details see
[\DMW]).  The configuration space here is the space of classical spins
which we describe by three-dimensional vectors of a given length (the
length ultimately gets related to the Casimir of the
$SU(2)$-representation).  This space is naturally embedded in ${\bf R}^3$;
we consider the latter  to be the dual space, $\Gamma$, to the Lie algebra
$su(2)$ under the following scalar product. Let $\{ x^i \}$ label the
points of ${\bf R}^3$ and let $\sum_{i=1}^3 a^i T^i$
denote the elements of $su(2)$ Lie algebra
where $T^i$ are generators of $su(2)$. We define a natural scalar
product between the two: $ \sum_i x^i a^i $.  Equivalently, in the matrix
notation $X \equiv
\sum_i x^i T^i$, $A\equiv \sum_i a^i T^i$,
we may write the scalar product as $\tr (XA) $.

The above scalar product has a natural interpretation in terms of
expectation value of the spin operator in a coherent state.
Consider a coherent
state of $SU(2)$, $| \vec x \ket$, belonging to the spin-$s$
representation, which satisfies the well-known property
$$ \bra \vec x| S^i | \vec x \ket = -s x^i. \eqn\threetwelvea$$
Here $S^i$ are the components of the quantum spin operator in the
spin-$s$ representation. The above scalar product can then be interpreted
as expectation value of the operator
$\sum_i a^i S^i$ in the coherent state $| \vec x\ket$.

Using the above scalar product one can define the coadjoint action of
$SU(2)$ on $\Gamma$. This simply  rotates the vector $\{ x^i\}$.
Hence, the coadjoint orbits of $SU(2)$ in ${\bf R}^3$ are spheres of
different radii. In terms of the matrix $X$ this means $X^2= $ constant.
We may now write down the action by Kirillov's construction:
$$ S[X] = i \int ds\, dt\, \tr ( X [ f_s, f_t]) + \int dt\, \tr(XB)
\eqn\fourthree$$
where  $f_t$ and $f_s$ are two tangent vectors on some given coadjoint
orbit at the point $X(t,s)$ and may be computed from
$$ \del_t X= [f_t, X],\;  \del_s X= [f_s, X].  \eqn\fourfour $$
The $f_t$ and $f_s$ are easily obtained from \fourfour\ using the
constraint $X^2 = $ constant. If we rescale $X$ for convenience to cast
the constraint in the form $X^2 = {\bf 1}$, then we find that
$$f_t = {1\over 4} [\del_t X, X], \; f_s = {1\over 4} [\del_s X, X]
\eqn\fourfoura  $$
so that the action can be rewritten as
$$ S[X] = {i\over 4} \int ds\, dt\, \tr (X[\del_t X, \del_s X]) + \int
dt\, \tr(XB),\quad  X^2 = {\bf 1}    \eqn\fourfive $$
Quantization is done by the path integral
$$Z \sim \int {\cal D} X(t) \prod_t \delta[X(t)^2 -1] \exp [i\lambda S[X]]
\eqn\fourfivea$$
where $\lambda$ is a constant. It is well-known that for the path-integral
to be well-defined $\lambda$ must be quantized. Different values of
$\lambda$ correspond to different spin representations. The theory then
knows about the underlying fermionic
structure by the specific choice of $\lambda$
corresponding to the spin-$1/2$ representation. Note that in the limit
$\lambda \to \infty$, the semiclassical method is exact.

The above procedure can be followed step-by-step for the present case of
\winf\ spin. The natural dual space, $\Gamma$, is the \winf-algebra
itself, which is the set of single-particle operators, or equivalently, is
the space of (generalized) functions on the phase space related in a
one-to-one fashion to the operators by the Weyl correspondence \twoten\
and \twoeleven.
Let us denote the points in $\Gamma$ by $\phi$.
Let us explicitly write it out in terms of the Heisenberg-Weyl basis
$$ \phi= \int d\alpha\, d\beta\, \tilde u(\alpha, \beta)\, \hat
g(\alpha, \beta) \eqn\foursevena$$
where
$$\tilde u(\alpha, \beta) = \int {dp \over 2\pi} {dq\over 2\pi}
e^{i(p\beta - q\alpha)} u (p,q)  \eqn\foursevenb$$
The analogy with the $SU(2)$ spin case is that $\hat g(\alpha, \beta)$ are
like the generators $T^i$, and $\tilde u(\alpha, \beta)$ (or $u(p,q)$) are
like the components $x^i$ of the point $\vec x$ in ${\bf R}^3$, and $\phi$
is like the matrix $X$. Moreover, $\phi$ can be interpreted in terms of
the expectation value of the bilocal operator $\Phi$ in a coherent state
of the $W_\infty$ algebra, just like the interpretation of $x^i$ as the
expectation value of the spin operator in an $SU(2)$-coherent state.

Just like in the $SU(2)$ case, there is a natural scalar product between
the points $\phi$ and elements  $\Theta$
(\twoten\ and \twoeleven) of $W_\infty$ Lie algebra. This scalar product is
$$ \bra \phi | \Theta \ket = \tr (\phi \Theta)   \eqn\fourfiveb$$
Under this scalar product, the coadjoint action on $\phi$ is defined in
the standard way. That is, corresponding to the infinitesimal
transformation  $\delta_\epsilon \Theta = {i\over \hbar}
[\epsilon, \Theta]$, $\phi$
transforms as  $\delta_\epsilon \phi = -{i\over \hbar}[\epsilon, \phi]$.
The compatibiliy of this coadjoint action with the scalar product is
obvious.

In terms of the phase space function $u(p,q)$ introduced in \foursevena\
and \foursevenb, the coadjoint action is easily deduced using \twoseven\
and is given by the Moyal bracket
\REF\MOYAL{J.E. Moyal, Proc. Cambridge Phil. Soc. 45 (1949) 99.}
[\MOYAL],
$$ \delta_\epsilon u(p,q)=  \{ \epsilon, u\}_{\sevenrm MB}(p,q)
\eqn\fourfivec$$
which  is defined  by
$$\{ A, B\}_{\sevenrm MB} (p,q) =
{2\over \hbar} \sin {\hbar \over 2}(\del_q
\del_{p'} - \del_p \del_{q'} ) [A(p,q) B(p',q')]_{p'=p, q'=q}
\eqn\fourfived$$
In the $\hbar \to 0$ limit it reduces to the Poisson bracket.

The specific coadjoint orbits of interest to us will be picked out by
imposing the constraints
$$\phi^2 = \phi, \quad \tr\, \phi = N  \eqn\fourfivee$$
in the dual space $\Gamma$. These constraints reflect an underlying
fermionic structure and can be understood as follows. The
``configuration'' $\phi$ is related to the fermion bilocal operator
$\Phi$ as
$$\phi = \bra \{ \phi\} | {\bf 1}-\Phi | \{\phi\}\ket
\eqn\fourfivef $$
where $| \{\phi\}\ket $ is a coherent state of $W_\infty$, analogous to
the state $| \vec x \ket$ in the $SU(2)$ case. The reason for the
appearance of ${\bf 1}- \Phi$ instead of just $\Phi$ can be traced to the
definition \threetwo, according to which it is the trace of ${\bf 1}-
\Phi$ which equals the number of fermions. This is also the origin of the
constraint $\tr\, \phi = N$ in \fourfivee. The origin of the other
constraint, $\phi^2=\phi$, can be traced to the operator constraint
\threetwelve, as can be seen by analyzing in detail its expectation value
in any coherent state. There is, however, a more direct way to see that
$\phi$ must satisfy the quadratic constraint.
Let us evaluate the expectation value
\fourfivef\  in the fermi ground state (which is a
coherent state in a trivial sense).
The corresponding configuration $\phi=\phi_0$ is given by
$$\phi_0 = \sum_{i\leq N} | i \ket \bra i | \eqn\fourfiveg$$
where $| i \ket, i=1,2,\cdots,\infty$
denote the energy eigenstates of the single-particle hilbert space.
Clearly $\phi_0$
satisfies the constraints \fourfivee. Moreover, it is clear from
\fourfivef\ that different configurations $\phi$ are related to each other
by similarity transformations  (i.e. by $W_\infty$-coadjoint
transformations). Therefore,
once we have shown that one point of the orbit satisfies
\fourfivee, we have proved it for the entire orbit.  From the form
\fourfiveg\ the fermionic character of our coadjoint orbit is clear.

Now that we know the dual space  and characterization of the coadjoint
orbits of interest, we can apply Kirillov's  method to construct the boson
action
$$ S[\phi,\a] = {i\over \hbar}
\int ds\, dt\, \tr ( \phi [ f_t, f_s]) - \int dt\, \tr(\phi\a)
\eqn\fourfiveh$$
where $f_t$ and $f_s$ are the hamiltonians on the coadjoint orbit that
lead to the motions
$$ i\hbar\del_t \phi= [f_t, \phi],\;
i\hbar\del_s \phi= [f_s, \phi].  \eqn\fourfivei $$
The crucial point is that using the constraint $\phi^2=\phi$ and
equations \fourfivei\ we can easily prove that
$$ S[\phi,\a] = {i\hbar} \int ds\,dt\;
\tr (\phi[\del_t \phi, \del_s \phi]) - \int
dt\, \tr(\phi\a) \eqn\fourfivek $$
This action can be written more explicitly in terms of the phase space
function $u(p,q,t,s)$ defined as in \foursevena\ and \foursevenb:
$$\eqalign{
S[u, \a]= \int ds\,dt
\int {dpdq\over 2\pi\hbar}& u(p,q,t,s)[\hbar^2\{ \del_s
u(p,q,t,s), \del_t u(p,q,t,s)\}_{\sevenrm MB}]\cr
+ \int dt&\int {dpdq\over 2\pi\hbar}h(p,q)u(p,q,t)\cr}
\eqn\foursevend$$
where $h(p,q)$ is the classical hamiltonian, $h(p,q) = \half (p^2 +
V(q))$, obtained from the background gauge field \twoonea, using the Weyl
correspondence. In terms of the $u$-variable the constraints read
$$\int {dpdq\over 2\pi\hbar} u(p,q) = N  \eqn\fourfivel$$
$$ \cos {\hbar \over 2} (\del_q \del_{p'} - \del_{q'} \del_p )
[u(p,q) u(p',q')]_{p'=p, q'=q} = u(p,q) \eqn\fourfivem$$

At this point  we wish to mention that in writing down the action
\fourfiveh\ and in the subsequent manipulations with it, we have made use
of the trace identity $\tr (\phi_1 \phi_2)= \tr (\phi_2 \phi_1)$. Since
$\phi$'s are infinite dimensional matrices, this identity is not satisfied
unless we put some restrictions on them. The simplest statement of the
restriction is
in terms of the corresponding phase space functions  in terms of which the
trace identity is equivalent to the condition
$\int {dp dq\over 2\pi \hbar} \{ u_1(p,q), u_2(p,q)\}_{\sevenrm MB} =0$.
Since the integrand can be written as a total derivative involving at
least one derivative on the $u$-function, we
can satisfy the above condition by
requiring $u(p,q) \to $ constant as $p,q\to \infty$.

\section{\bf Classical equation of motion and its solutions:}

The most general variation of $\phi$, consistent with the
constraints $\phi^2= \phi$ and $\tr\, \phi=N$, is
$$ \phi \rightarrow V\phi V^\dagger, \qquad VV^\dagger={\bf 1}.
\eqn\fiveone$$
That is, the independent variables are the \winf\ ``angles".
To obtain the
classical equation of motion from the action \foursevend, therefore, we
make the above variation \fiveone\ in $\phi$, with $V = 1+ i\Theta$,
$\Theta$ infinitesimal. The change in the action is
$$ \delta S[\phi, \a] = -\hbar \int ds\, dt\, [\del_s \{\tr (\Theta \del_t
\phi)\} - \del_t\{ \tr(\Theta \del_s \phi)\}] + i\int dt\, \tr(\Theta [\a,
\phi])   \eqn\fivetwo$$
We shall take time $t$ to be non-compact. Then the $(s,t)$ space is a
half plane, with $-\infty \leq s \leq 0,\, -\infty \leq t \leq + \infty$
and the boundary conditions $\phi(t,s=-\infty)= {\bf 1}$ and $\phi(t,s=0)=
\phi(t)$. Also, assuming that $\phi(t) \to {\bf 1}$ as $t\to \pm \infty$,
only the $s$-boundary term contributes in \fivetwo\ and we get
$$ \delta S = - \int dt\, \tr [ \Theta ( \hbar \del_t \phi - i[\a,
\phi])].  \eqn\fivethree$$
This gives the equation of motion
$$ i\hbar \del_t \phi+[\a, \phi]=0, \quad \phi^2= \phi,\;\tr\phi=N
\eqn\fivefour$$
Classically, therefore, the \winf\ spin system under consideration is
completely defined by \fivefour. We will
now solve this equation and show that
the constraints $\phi^2= \phi,\; \tr\,\phi=N$ keep track
of the underlying fermionic  structure.

Expanding $\a$ and $\phi$ in the Heisenberg-Weyl basis, we may write
$$ \a = \int d\alpha \, d\beta\, [\half (\del_\beta^2 - \del_\alpha^2
+ i{g_3\over \sqrt N}\del_\alpha^3 + \cdots)
\delta (\alpha) \delta(\beta)] \hat g(\alpha, \beta)   \eqn\fivenineteen$$
and \foursevena\ and \foursevenb\ for $\phi$.
The equation of motion for $u(p,q,t)$ now becomes
$$ \del_t u= \{ h, u\}_{\sevenrm MB}   \eqn\fivetwenty$$
where $$h(p,q)= \half(p^2 - q^2 + {g_3 \over \sqrt N} q^3 + \cdots).
\eqn\oldzero$$

\underbar{Time-independent case}:

In this case the equation of motion is solved by any $u$
which depends on $p,q$ only through the  function $h(p,q)$. That is to say,
in the phase space $u$ takes the same value on curves of constant
classical energy. Out of all such $u$'s, the classical problem is solved
only by those that satisfy the quadratic constraint \fourfivem. This
constraint cannot be solved exactly, except in the limit $\hbar \to 0$.
Denoting $u$ by $u^{(0)}$ in this limit, \fourfivem\ leads to
$$ (u^{(0)}(p,q))^2 = u^{(0)}(p,q) \eqn\fivetwentyfive$$
Thus $u^{(0)}$ takes the same value (1 or 0)
on curves of constant energy
in phase space. For example, one may choose
$$ u^{(0)}(p,q) = \theta (\e_F - h(p,q)). \eqn\fivetwentysix$$
$\e_F$ is a parameter of this classical solution.
Finally $u^{(0)}(p,q)$ must
satisfy the fermion number constraint \fourfivel;
for a $u^{(0)}$ of the above form this fixes $\e_F$ in terms of $\hbar$
and $N$. For the hamiltonian \oldzero\ we find\ $-\e_F\sim 1/(\hbar N)$
which is consistent with the fact that in the double scaling limit we
treat $-\e_F N \equiv \mu$ as the inverse string coupling.
It is clear that \fivetwentysix\ is
just the classical phase space density of fermions in the fermi
vacuum. We have thus once again arrived at the underlying fermionic
picture.

\underbar{Time-dependent case}:

As in the time-independent case we are able to solve the equations only in
the $\hbar\to 0$ limit. Let us denote the solution of the constraint in
this limit by the characterisitic function $\chi_{R(t)}(p,q)$, which
satisfies \fivetwentyfive\ and defines a region $R(t)$ of phase space.
There is a time-dependence in $R(t)$ because in the present case the
region changes with time. The region $R(0)$
can in principle be quite complicated involving several fluid blobs or
droplets of the fermi fluid. Since we are working in the $\hbar \to 0$
limit, the equation of motion satisfied by $u(p,q,t)$ reduces to the
classical one: $\del_t u=\{h(p,q), u(p,q)\}_{\sevenrm PB}$. It can be
easily shown that $u= \chi_{R(t)}$ satisfies the equation of motion if
the region $R(t)$ is given by
$$\chi_{R(t)}(p,q)= \chi_{R(0)}(\bar p(t), \bar q(t)) \eqn\oldone$$
where $(\bar p(t),\bar q(t))$  denote the classical trajectory  evolving
according to the  hamiltonian
$-h(p,q)$ with the initial conditions
$\bar p(t=0)=p,\; \bar q(t=0)=q$. In other
words, the region $R(t)$ is obtained by evolving each point in the region
$R(0)$ for time $t$ under the classical hamiltonian $h$.
For the hamiltonian \oldzero\
we can write the classical trajectories explicitly
if we ignore the $O(1/\sqrt N)$ terms. This leads to
$$\chi_{R(t)}(p,q)= \chi_{R(0)}(p\cosh t- q\sinh t,-p\sinh t + q\cosh t).
\eqn\oldtwo$$

\section{\bf Correspondence with Geometric Action for Fluid Profiles}

In this section we would like to show how the geometric action for fluid
profiles [\DMW] may be obtained from the exact classical action in the
limit $\hbar \to 0$.  As we mentioned in the last section the
characteristic function of a region $R$ in phase space satisfies the
constraint \fourfivem\ in the limit $\hbar \to 0$. This reflects the fact
that as $\hbar\to 0$
the phase space density $u(p,q)$ corresponds to that of
an incompressible fermi fluid whose density is $1$ in some region $R$ and
$0$ outside. When we consider a two-parameter deformation (in $(t,s)$) of
the phase space density $u(p,q,t,s)$, classically it corresponds to a
two-parameter deformation $R(t,s)$ of the fluid region
$R$. In the following  we shall therefore put
$$ u(p,q,t,s) = \chi_{R(t,s)}(p,q) + \hbar \,{\rm
corrections}.
\eqn\newzero$$

The correspondence with the fluid-profile action [\DMW] is most directly
made by rewriting the action \fourfiveh\ and the equations \fourfivei\ in
terms of the phase space variables. Let us denote the first term of
\fourfiveh\ by $S_0$. In terms of the phase space variables it reads
$$S_0= \int ds\, dt\, \int {dpdq\over 2\pi\hbar} u(p,q,t,s) \{f_t,
f_s\}_{\sevenrm MB}  \eqn\newzeroa$$
The hamiltonians $f_t$ and $f_s$ are defined by \fourfivei. In terms of
phase space variables equations \fourfivei\ read
$$\del_t u= \{f_t, u\}_{\sevenrm MB},\quad
\del_s u= \{f_s, u\}_{\sevenrm MB} \eqn\newzerob$$
In the limit $\hbar\to 0$, the Moyal bracket goes over to the Poisson
bracket. Using this fact and equation \newzero\ we get
$$S_0= \int ds\, dt\, \int {dpdq\over 2\pi\hbar} [\chi_{R(t,s)} \{f_t,
f_s\}_{\sevenrm PB} + o(\hbar^2)] \eqn\newzeroc$$
where
$$\del_t \chi_R= \{f_t, \chi_R\}_{\sevenrm PB},\quad
\del_s \chi_R= \{f_s, \chi_R\}_{\sevenrm PB} \eqn\newzerod$$
It is simple to see that \newzeroc\
is the same as the action $S_0$ written
in equation (62) of [\DMW]. To facilitate the comparison, let us recall
that equation (62) of [\DMW] is
$$ S_0= \int dt\, ds\,\bra \chi_{R(t,s)}(p,q)\,|\,
[\del_t U U^{-1}, \del_s U U^{-1} ]\ket \eqn\newtwo$$
which is equivalent to
$$ S_0= \int dt\, ds \int {dpdq\over 2\pi \hbar}
\chi_{R(t,s)}(p,q) \{f_t,f_s\}_{\sevenrm PB}
 \eqn\newthree$$
where we have written out the definition of the scalar product used in the
last paper, and used the fact that $\del_a U U^{-1}, a=s,t$ are
Lie algebra elements  corresponding to the functions $f_s, f_t$ satisfying
the property \newzerod\
(we have explained in [\DMW] how the commutatator in
the $w_\infty$ Lie algebra is equivalent to Poisson bracket of functions
in phase space).

The second term in \fourfiveh\ in the limit $\hbar\to 0$ becomes the
classical energy contained in the fluid region $R(t,s)$, which is the same
as equation (66) of [\DMW]. Therefore we see that the action written in
the present paper agrees with the one in [\DMW] in the limit
$\hbar\to 0$.  A different approach to the classical limit is discussed in
\REF\IKS{S. Iso, D. Karabali and B. Sakita, City College
preprint, 1992.}
[\IKS]. A different approach to coadjoint orbits of $w_\infty$, the group
of area-preserving diffeomorphisms, is discussed in
\REF\AJ{J. Avan and A. Jevicki, Mod. Phys. Lett. A7 (1992) 357.}
[\AJ].

\section{\bf Interacting Fermions}

In this section we indicate how the bosonization technique described so far
can be applied to a wide class of
interacting fermi systems.

Let us try to use the bilocal operator $\phi$ or equivalently the phase
space density $u(p,q)$ again as the basic dynamical variable.
The first point to realize is
that the constraints \fourfivee
(equivalently \fourfivel\ and \fourfivem) have been derived above
using purely kinematic reasoning without considering the equation of
motion of the fermi field. This is obvious for \fourfivel, which
simply states that the total number of fermions is $N$, a condition that
is satisfied by any closed system of fermions, interacting or otherwise.
The second constraint, \fourfivem, originates from the  operator
constraint \threetwelve, {\it viz.}, $\Phi^2 = (N+1) \Phi$. The only
ingredient that went into the derivation of this constraint is the
anticommutation relation of the fermi field which again does not depend on
the dynamics of the fermi system.
Besides the constraints, the first
term (the symplectic form) in the action \foursevend\ is also
purely kinematic, and it does not depend on the choice of the
many-body hamiltonian. With these remarks, it is now easy to see that the
coadjoint orbits of \winf\ that we have constructed are suitable for
representing interacting fermions also, provided
the interaction
can be expressed in terms of $\phi$ or equivalently $u(p,q)$.

Let us now give some examples. The most general interaction involving
quadratics of $\phi$ (or $u$) is
$$S_{\rm int}= \int dt \int dx\,dy\,dz\,dw\, {\cal A}_{xyzw}
\phi(x,y) \phi(z,w)  \eqn\sevenone$$
The $SU(2)$-spin analog of such a term would  be $\sum_{ij}
S_i S_j {\cal B}_{ij}$ which can be thought of as coupling to some
``generalized  magnetic fields'' which have tensorial transformation
properties under $SU(2)$ rotations (instead of vector transformations
which are true of usual magnetic fields).
The above interaction term would be the bosonized form of the following
fermion interaction:
$$S_{\rm int}= \int dt \int dx\, dy\, dz\, dw\, {\cal A}_{xyzw}
\psi(x,t) \psi^\dagger(y,t) \psi(z,t) \psi^\dagger(w,t) \eqn\seventwo$$
Clearly, the standard four-fermi interaction $\int dt\, dx\,
[\psi^\dagger(x)\psi(x)]^2$
is included in this list.

The generalization to cubic and higher interaction terms in $\phi$ is
obvious; they just involve introduction of higher external
tensor fields of \winf\
in the sense explained above. It would be extremely interesting to
understand emergence of new collective excitations like plasmons arising
out of interacting bose theories such as the ones mentioned above.

\section{\bf Concluding Remarks}

In this paper we have presented solution of the bosonization problem of
non-relativistic fermions in $1$-dimension. We believe that this
formulation will give us a handle on some important issues of
two-dimensional string theory.
For instance, using our action \fourfived\
and the constraints \fourfivel\
and \fourfivem, we can look for  stringy non-perturbative
effects $\sim \exp(-1/\hbar)$
that have been discussed by Shenker
\REF\SHENKER{S. Shenker, in Proceedings of the Cargese Workshop on Random
Surfaces, Quantum Gravity and Strings, 1990, edited by O. Alvarez, E.
Marinari and P. Windey (Plenum Press).}
[\SHENKER]. We should note in this context that both our classical
action and the constraint \fourfivem\
contain explicit factors of $\hbar$.
The second point is that we have seen that a $(1+1)$ dimensional
target space picture emerges from the $c=1$ matrix model perturbatively in
$\hbar$, the string coupling constant.
Since our formulation is valid for
all values of $\hbar$ it is clearly important to ask what
happens to this picture for large $\hbar$, i.e. in the strong coupling
limit. We have not been able to solve the constraint
\fourfivem\ in this limit, but it seems to us that the above picture
of a $(1+1)$-dimensional target space theory cannot be valid in this
limit. The approximation to a single fluid blob, valid for small $\hbar$,
must necessarily break down as the string coupling constant increases,
which is also accompanied with the loss of incompressiblity of the fermi
fluid on account of large quantum corrections to step-function-like
densities.  One presumably then has numerous fluid blobs all over the
phase space indicating that in the limit $\hbar \to \infty$, one may
have to deal with the full $(2+1)$-dimensional theory described by the
action \foursevend.
Such a scenario implies that the standard reasoning from the viewpoint of
continuum quantum gravity that the dynamical metric is equivalent to one
conformal mode (Liouville)
which in turn is equivalent to \underbar{one} additional target space
dimension, breaks down in the strong coupling limit.
It is clearly very important to make this discussion
more concrete.

Another interesting aspect of the strong coupling limit is the following.
{}From the viewpoint of fermions moving in the inverted harmonic oscillator,
$\hbar \to \infty$ limit is equivalent to $\mu \equiv -N\e_F \to 0^+$
(we are measuring $\e_F$ with respect to the top of the potential
and using the convention that $\mu$ is positive for energies below the
top).
In this limit the fermi level moves to  the top of the potential. Since
the potential barrier is negligible here, the fluid  will freely move
between the two ``classical worlds'' described by the inverted harmonic
oscillator potential.
In order to gain some insight into the description of the $\mu\to 0^+$
limit, it may be useful to consider a generalized model in which we
consider the entire range $\mu\in (-\infty, +\infty)$. For $\mu \to 0^-$,
this model does not correspond to a string theory, in the sense that the
perturbation expansion of the matrix model fails to exist. However,
negative $\mu$'s make perfect sense as a theory of fermions. The
generalized model has another classical limit as $\mu \to -\infty$
in addition to the weak coupling string theory ($\mu \to \infty$).
It would be interesting to see if the two different signs of $\mu$
correspond to two different phases of the fermi system.
Such phase transitions are known to
occur in $2+1$ dimensional fermi systems in the disussion of
quantum Hall effect
\foot{We thank F.D.M. Haldane for  pointing this out to us.}.
We have also seen that our bosonization techniques can be applied to
interacting fermion systems in one dimension.
It would
be worthwhile if some of the techniques introduced in this paper can
deepen our understanding of $(1+1)$-  and  $(2+1)$- dimensional condensed
matter systems.

\noindent{\bf Acknowledgements}:
A.D. would like to thank International Centre  for Theoretical Physics,
Trieste and Niels Bohr Institute for hospitality during a visit during
which part of this work was done.
S.R.W would like to thank International Centre  for Theoretical Physics,
Trieste and CERN for hospitality during a visit during which part of the
work was done. S.R.W. would like to thank L.
Alvarez-Gaume, Caeser Gomez and F.D.M. Haldane for useful comments on the
connection of this work with the quantum Hall effect.

\refout
\end